\documentclass[aps,prb,twocolumn,superscriptaddress]{revtex4}
\usepackage{siunitx}
\usepackage{wrapfig}
\usepackage{amsmath}
\usepackage{amssymb}
\usepackage{graphicx}
\usepackage{natbib}

\begin{document}

\title{Selection rules in a strongly coupled qubit-resonator system}
\author{T.~Niemczyk}
\affiliation{Walther-Mei{\ss}ner-Institut, Bayerische Akademie der
Wissenschaften, D-85748~Garching, Germany}
\author{F.~Deppe}
\affiliation{Walther-Mei{\ss}ner-Institut, Bayerische Akademie der
Wissenschaften, D-85748~Garching, Germany}
\affiliation{Physik-Department, Technische Universit\"{a}t
M\"{u}nchen, D-85748 Garching, Germany}
\author{E.~P.~Menzel}
\affiliation{Walther-Mei{\ss}ner-Institut, Bayerische Akademie der
Wissenschaften, D-85748~Garching, Germany}
\author{M.~J.~Schwarz}
\affiliation{Walther-Mei{\ss}ner-Institut, Bayerische Akademie der
Wissenschaften, D-85748~Garching, Germany}
\author{H.~Huebl}
\affiliation{Walther-Mei{\ss}ner-Institut, Bayerische Akademie der
Wissenschaften, D-85748~Garching, Germany}
\author{F.~Hocke}
\affiliation{Walther-Mei{\ss}ner-Institut, Bayerische Akademie der
Wissenschaften, D-85748~Garching, Germany}
\author{M.~H\"{a}berlein}
\affiliation{Walther-Mei{\ss}ner-Institut, Bayerische Akademie der
Wissenschaften, D-85748~Garching, Germany}
\author{M.~Danner}
\affiliation{Walther-Mei{\ss}ner-Institut, Bayerische Akademie der
Wissenschaften, D-85748~Garching, Germany}
\author{E.~Hoffmann}
\affiliation{Walther-Mei{\ss}ner-Institut, Bayerische Akademie der
Wissenschaften, D-85748~Garching, Germany}
\author{A.~Baust}
\affiliation{Walther-Mei{\ss}ner-Institut, Bayerische Akademie der
Wissenschaften, D-85748~Garching, Germany}
\author{E.~Solano}
\affiliation{Departamento de Qu\'{\i}mica F\'{\i}sica, Universidad
del Pa\'{\i}s Vasco - Euskal Herriko Unibertsitatea, Apdo. 644,
48080 Bilbao, Spain} \affiliation{IKERBASQUE, Basque Foundation
for Science, Alameda Urquijo 36, 48011 Bilbao, Spain}
\author{J.~J.~Garcia-Ripoll}
\affiliation{Instituto de F\'{i}sica Fundamental, CSIC, Serrano
113-bis, 28006 Madrid, Spain}
\author{A.~Marx}
\affiliation{Walther-Mei{\ss}ner-Institut, Bayerische Akademie der
Wissenschaften, D-85748~Garching, Germany}
\author{R.~Gross}
\affiliation{Walther-Mei{\ss}ner-Institut, Bayerische Akademie der
Wissenschaften, D-85748~Garching, Germany}
\affiliation{Physik-Department, Technische Universit\"{a}t
M\"{u}nchen, D-85748 Garching, Germany}

\date{\today}

\begin{abstract}
\noindent Superconducting qubits acting as artificial two-level atoms allow for controlled variation of the symmetry properties which govern the selection rules for single and multiphoton excitation. We spectroscopically analyze a superconducting qubit-resonator system in the strong coupling regime under one- and two-photon driving. Our results provide clear experimental evidence for the controlled transition from an operating point governed by dipolar selection rules to a regime where one- and two-photon excitations of the artificial atom coexist. We find that the vacuum coupling between qubit and resonator can be
straightforwardly extracted from the two-photon spectra where the detuned two-photon drive does not populate the relevant resonator mode significantly.
\end{abstract}
\maketitle

\section{Introduction\label{sec:introduction}}

\noindent Superconducting quantum circuits have proven to be versatile model systems for performing quantum optics experiments on a chip, allowing for the investigation of fundamental quantum mechanics and scenarios for scalable quantum information
processing. With respect to the former, circuit quantum electrodynamics (QED)~\cite{Wallraff:2004a,Schuster:2007a,Houck:2007a,Astafiev:2007a,Abdumalikov:2008,Hofheinz:2008a,Niemczyk:2010}
has gained exceptional importance in recent years. Here, the coupling of natural atoms to the photon field inside a three-dimensional optical cavity, known as cavity QED,~\cite{Haroche:2006a} is modeled using superconducting circuits. In particular,
suitably designed Josephson junction based quantum circuits act as artificial two-level atoms (qubits) and transmission line resonators serve as ``boxes'' for microwave photons. Therefore, circuit QED systems are extremely attractive because
high qubit-resonator coupling strengths can be attained due to huge effective dipole moments of the artificial atoms and the small mode volumes of the resonators. For this reason, the strong coupling regime, where qubit and resonator exchange photons coherently, can be realized easily despite the
limited coherence times typically ranging from a few nano- to a few microseconds. Recently, even the new regime of ultrastrong light matter coupling has been reached,~\cite{Niemczyk:2010} where the coupling strength becomes a significant fraction of the total system energy. Another important key feature of circuit
QED is the design flexibility inherent to circuits on a chip and the ability to tune their properties \emph{in situ} over a wide range of parameters. This has triggered experimental studies of multi-photon driven artificial atoms,
including population inversion,~\cite{Ilichev:2010a} Mach-Zehnder interferometry,~\cite{Oliver:2005a} Landau-Zener
interference,~\cite{Sillanpaa:2006a} qubits coupled to microscopic defects,~\cite{Lupascu:2009a} amplitude spectroscopy,\cite{Berns:2005} multi-photon spectroscopy of hybrid quantum systems,\cite{Bushev:2010} and
sideband transitions.~\cite{Wallraff:2007a} Recently, an experimental investigation of a two-photon driven flux qubit coupled to a lumped-element $LC$ resonator demonstrated the controllability of fundamental symmetry
properties of circuit QED systems.\cite{Deppe:2008a,Niemczyk:2009a} However, in
that work, a high loss rate of the $LC$ resonator leading to a weakly coupled qubit-resonator system in combination with the presence of microscopic defects complicated the analysis of the data. Here, we report on one- and two-photon spectroscopy of a qubit-resonator system in the strong coupling limit. Our
results provide clear experimental evidence for the controlled transition from an operating point governed by dipolar selection rules~\cite{Liu:2005} to a regime where one- and two-photon excitations of the artificial atom coexist.
While the former is strictly analogous to the case  of natural atoms, the latter regime is a particular prominent feature of circuit QED. In addition, we find that the vacuum coupling strength between qubit and resonator can be straightforwardly extracted from the two-photon spectra since in this case the
detuned two-photon drive does not populate the relevant resonator mode significantly.

\noindent The paper is structured as follows: in Sec.~\ref{sec:setup}, we introduce the sample design, fabrication details, measurement setup and techniques. In Sec.~\ref{sec:spectroscopy}, we present and discuss the results of one- and
two-photon spectroscopy of the strongly coupled qubit-resonator system. In Sec.~\ref{sec:symmetry} we discuss the symmetry properties of our artificial atom and their implications for multi-photon driving in circuit QED. Finally, Sec.~\ref{sec:conclusions} is dedicated to concluding remarks.

\begin{figure}[tb]
\includegraphics[width=0.9\linewidth]{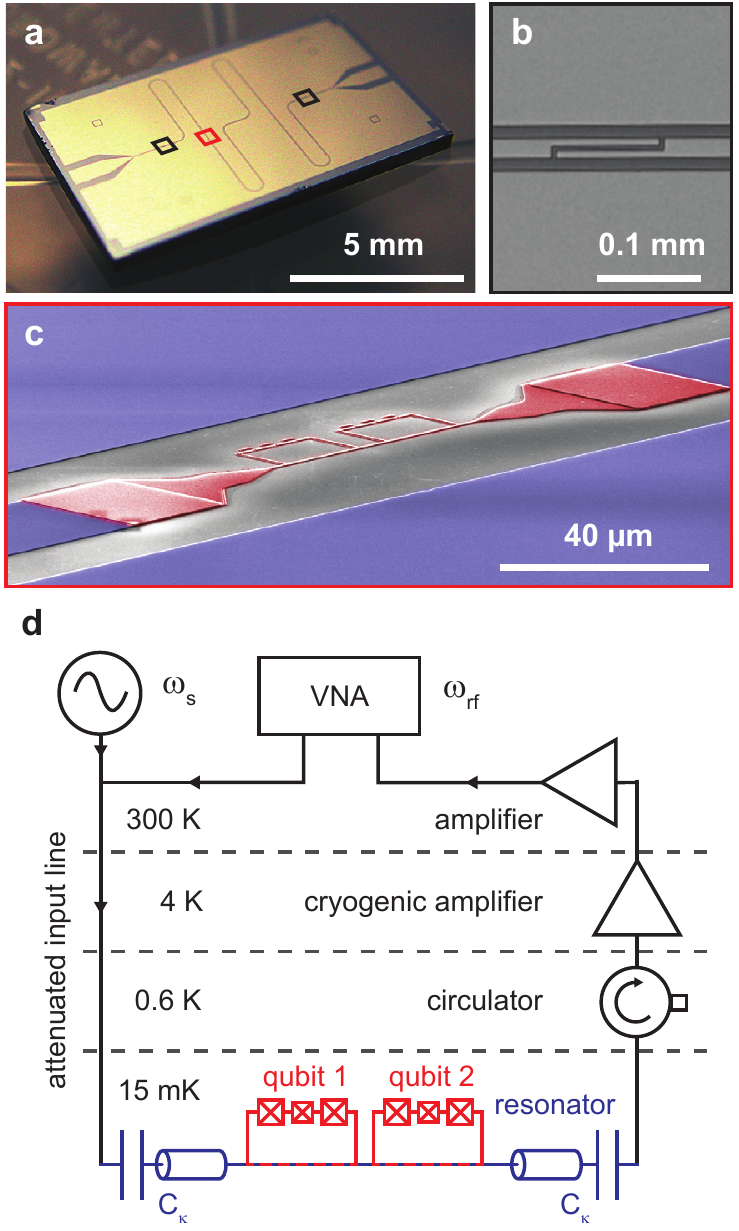}
\caption{(Color online) (a) Photograph of the superconducting coplanar
waveguide (CPW) resonator. The position of the qubit is indicated by the red box while the positions of the
in- and output capacitors are indicated by black boxes. (b) SEM micrograph of a coupling capacitor with an estimated capacitance $C_{\kappa}\approx\SI{10}{\femto\farad}$. The two fingers are separated by a gap of \SI{4}{\micro\metre} and are approximately \SI{100}{\micro\metre} long. (c) False-color SEM micrograph of the two flux qubits (red) galvanically connected to the center conductor of a niobium CPW resonator (blue) fabricated on a silicon substrate (gray). The qubit loop areas are approximately $8.5$~x~$14~\mu\text{m}^2$. (d) Schematic of the measurement setup. The coupled qubit-cavity system is probed through a highly attenuated input line using a vector network analyzer (VNA). The transmitted signal is amplified at \SI{4}{\kelvin} with a cryogenic HEMT amplifier. A second microwave source
excites the system in spectroscopy experiments.}\label{FIG1}
\end{figure}

\section{Setup and experimental techniques\label{sec:setup}}

\noindent Figure~\ref{FIG1} shows our quantum circuit and a schematic of the measurement setup. In our experiments, a superconducting persistent flux qubit~\cite{Mooij:1999a,Deppe:2007a} is
galvanically~\cite{Abdumalikov:2008,Niemczyk:2010} connected to the center conductor of a niobium coplanar waveguide resonator~(Fig.\,\ref{FIG1}\,a-c). The niobium film with a thickness of \SI{100}{\nano\metre}
is deposited by dc-magnetron sputtering on a thermally oxidized (\SI{50}{\nano\metre}) silicon substrate. The patterning of the Nb films is done by optical lithography and reactive ion etching. The center conductor has a width of \SI{20}{\micro\metre} and is separated from the lateral ground
planes by a \SI{12}{\micro\metre} gap, leading to a characteristic impedance of about \SI{50}{\ohm}. The high-$Q$ resonant cavity ($3500 \lesssim Q \lesssim 6300$; depending
on the mode frequency) is coupled via small gap capacitors $C_{\kappa}$ to the readout circuitry and has a fundamental resonance frequency ($\lambda/2$-mode) of $\omega_1/2\pi=\SI{2.745}{\giga\hertz}$. At a current antinode of the $\lambda$-mode ($\omega_2/2\pi=\SI{5.324}{\giga\hertz}$), a \SI{80}{\micro\metre} section of the center conductor is replaced by a narrow (\SI{500}{\nano\metre}) aluminum strip. Two flux qubits are galvanically coupled to this narrow strip (see Fig.~\ref{FIG1}\,c). The inhomogeneous geometry of the transmission line leads to a position dependent inductance
$L(x)$ and capacitance $C(x)$ per unit length which in turn results in a non-harmonic distribution of the higher mode frequencies, $\omega_{j} \neq j\cdot \omega_{1}$ with $j \in \mathbb{N}$. Each flux qubit consists of three Al/AlO$_{\rm x}$/Al Josephson junctions (JJ) interrupting a superconducting loop. The qubits and
the narrow aluminum strip are fabricated by electron beam lithography and shadow evaporation technique. The bottom and top aluminum layer have a thickness of \SI{50}{\nano\metre} and \SI{70}{\nano\metre}, respectively. Two of the JJ have nominal areas $A \approx 0.03~\mu\text{m}^2$, while one
junction is smaller by a factor of $\alpha \approx 0.63$. A schematic of the measurement setup is shown in Fig.~\ref{FIG1}\,d. The sample is placed at the
base temperature of \SI{15}{\milli\kelvin} in a dilution refrigerator. We monitor the amplified cavity transmission signal at $\omega_{\rm rf}$ of the coupled qubit-resonator system using a vector network
analyzer (VNA). For spectroscopy measurements, the qubit can be excited by an additional microwave tone $\omega_{\rm s}$ with power $P_{\rm s}$. A cryogenic circulator prevents amplifier noise from entering our experiment.\\

\noindent The potential of the three-junction flux qubit can be reduced to a one-dimensional double well potential, where the two minima correspond to opposite persistent currents $\pm I_{\rm p}$. This double well potential can be tuned by an external flux bias $\Phi_{\rm x}$. At $\delta\Phi_{\rm x} = \Phi_{\rm x} - (k - 0.5)\Phi_{\rm 0} = 0$, these current states are degenerate. Here, $\Phi_{\rm 0} = h/2e$ denotes the magnetic flux quantum and $k \in \mathbb Z$. The finite tunnel coupling between these states results in a symmetric and antisymmetric superposition state separated by an energy gap $\Delta$. The two flux qubits are fabricated with slightly different loop areas by varying the electron exposure dose during electron beam lithography. For $|k| \geq 2$, the small area difference allows to investigate the individual qubit parameters separately by making use of the $\Phi_{\rm
0}$-periodicity of the qubit potential. The spectroscopy measurements presented in this work are conducted around $\Phi_{\rm x} \approx -2.5~\Phi_{\rm 0}$ ($k=-2$), where the qubit transmission spectra are well separated in the
relevant experimental frequency regime. In the remainder of this paper we solely focus on one of the qubits,\cite{Remark:1} which we refer to as qubit 1. The qubit-resonator system can be described quantum mechanically by a multi-mode Jaynes-Cummings Hamiltonian
\begin{equation}
\hat{H}  =  \hbar\omega_{\rm q}\hat{\sigma}_{\rm z}/2  +  \sum\limits_{j}
\left[\hbar\omega_{j}\hat{a}_{j}^{\dagger}\hat{a}_{j} + \hbar \tilde{g}_{j}\left(\hat{a}_{j}^{\dagger}\hat{\sigma}_{-}
+ \hat{a}_{j}\hat{\sigma}_{+} \right)\right].\label{FullH}
\end{equation}
Here, $\omega_{\rm q}=\sqrt{\Delta^2 + (2I_{\rm p}\cdot\delta\Phi_{\rm x})^2}/\hbar$ is the flux dependent qubit transition frequency and $\hat{\sigma}_{\rm z}$ a Pauli spin operator. The operator $\hat{a}_{j}^{\dagger}$ ($\hat{a}_{j}$) creates (annihilates) a photon in the
resonator mode with frequency $\omega_{j}$ while $\hat{\sigma}_{\rm +}$ ($\hat{\sigma}_{\rm -}$) creates (annihilates) an excitation in the qubit. The qubit-resonator coupling strength $\tilde{g}_{j} = g_{j}\sin\theta$ is flux
dependent and exhibits a maximum at $\delta\Phi_{\rm x} = 0$
\begin{figure}[tb]
\includegraphics[width=0.9\linewidth]{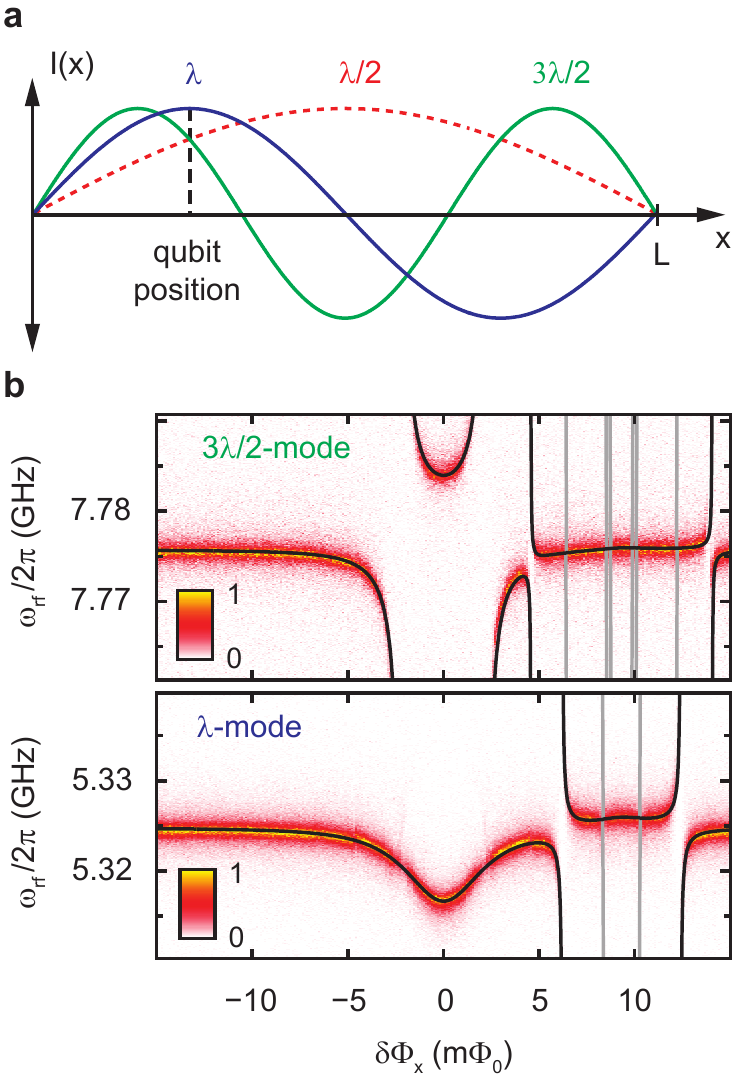}
\caption{(Color online) (a) Sketch of the current distribution of the first three modes of the coplanar waveguide resonator (CPW) with length $L = \SI{23}{\milli\metre}$. The resonance frequencies at $\Phi_{\rm x}= 0$ are: $\omega_{\rm 1}/2\pi = \SI{2.745}{\giga\hertz}$ ($\lambda/2$-mode, dashed red), $\omega_{\rm 2}/2\pi =
\SI{5.324}{\giga\hertz}$ ($\lambda$-mode, blue) and $\omega_{\rm 3}/2\pi = \SI{7.775}{\giga\hertz}$ ($3\lambda/2$-mode, green). The current distribution for the $\lambda/2$-mode is plotted dashed as its presence can be safely neglected for the qubit under consideration. (b) Cavity transmission spectrum (linear scale, arb.\,units) for the $3\lambda/2$-mode (top panel) and the $\lambda$-mode (bottom panel) as a function of the relative flux bias $\delta\Phi_{\rm x}$ of qubit 1 and the probe frequency $\omega_{\rm rf}$. The spectra are recorded at a flux bias of $\Phi_{\rm x} \approx -2.5~\Phi_{\rm 0}$, where the two qubits present in the system can be studied independently. The solid lines show the numerically evaluated full energy level spectrum, where we included the second qubit into Hamiltonian (\ref{FullH}) to verify our assumption of negligible qubit-qubit coupling.\cite{Remark:1} In contrast to the black lines, the gray lines represent energy levels of states with more than one excitation.\cite{Niemczyk:2010}}\label{FIG2}
\end{figure}
\noindent where $\sin\theta = \Delta/\hbar\omega_{\rm q}=1$. The maximum qubit-resonator coupling strength can be expressed as $\hbar g_{j} = M I_{\rm p} I_{j}$, where $M$ denotes the
mutual inductance between qubit and cavity and $I_{j}$ is the vacuum current of the $j$th resonator mode. Owing to our sample design, $g_{j}$ is maximum for the $\lambda$-mode (see Fig.~\ref{FIG2}\,a). In the sum of the Hamiltonian~(\ref{FullH}) we only have to include the experimentally relevant cavity modes ($j=2,3$; $\lambda$- and $3\lambda/2$-mode, blue and green line in
Fig.~\ref{FIG2}\,a, respectively). For the qubit under consideration, it is perfectly fine to neglect the presence of the $\lambda$/2-mode ($j=1$; dashed red line in Fig.~\ref{FIG2}\,a) in the theoretical modeling as it is both largely detuned from $\omega_{\rm q}$ and only weakly coupled. The same is true for the $j \ge 4$ modes. Figure~\ref{FIG2}\,b shows the cavity transmission spectra for the $\lambda$-and $3\lambda/2$-mode as a function of $\delta\Phi_{\rm x}$. The transmission
spectrum of the $3\lambda/2$-mode (Fig.~\ref{FIG2}\,b, top panel) reveals avoided crossings at flux positions ($\delta\Phi_{\rm x} \approx \pm 2~\text{m}\Phi_{\rm 0}$) where the transition frequency of the qubit $\omega_{\rm q}$ equals the resonator mode frequency $\omega_{\rm 3}$. At these
flux values, the system is resonant and energy can be coherently exchanged between qubit and resonator. There are no avoided crossings in the transmission spectrum of the $\lambda$-mode for qubit 1 (Fig.\,\ref{FIG2}\,b, bottom panel). Here, the system is in the dispersive regime for all $\delta\Phi_{\rm x}$ values. In this regime the detuning $\delta_{j} = |\omega_{\rm q} - \omega_{j}|$ between the qubit and the $j$th cavity mode is large in comparison to $g_{j}$. Although qubit and resonator can not exchange energy directly,
they are still interacting through a second order dispersive coupling. Therefore, we observe an ac-Zeeman shift of the cavity's resonance frequency $\omega_{\rm 2}$ which depends on the qubit state. Thus the ac-Zeeman shift can be utilized for two-tone spectroscopy measurements.~\cite{Schuster:2005a,Abdumalikov:2008} The measurement protocol for this spectroscopy technique is as follows: for fixed $\delta\Phi_{\rm x}$, we monitor both phase and amplitude at the probe tone frequency $\omega_{\rm
rf}$ where maximum cavity transmission occurs. A second microwave tone, the spectroscopy tone ($\omega_{\rm s}$), is applied to the system. When $\omega_{\rm s} \approx \omega_{\rm q}$ the qubit is excited and for sufficiently large amplitude of the continuous spectroscopy tone the transition is saturated, yielding an equal population of ground and excited state. This leads to a shift of the cavity's resonance frequency by $g_{j}^2/\delta_{j}$ which results in a decrease in magnitude and a phase shift of the transmitted
signal at $\omega_{\rm rf}$. Furthermore, the dispersive qubit-cavity interaction modifies the qubit's transition frequency which is then given by $\tilde{\omega}_{\rm q} = \omega_{\rm q} + 2g_{j}^2\hat{a}_{j}^{\dagger}\hat{a}_{j}/\delta_{j} + g_{j}^2/\delta_{j}$.\cite{Schuster:2005a} The last term represents the Lamb shift due to the presence of the vacuum. The ac-Zeeman shift proportional to $\bar{n}_{j} =
\langle\hat{a}_{j}^{\dagger}\hat{a}_{j}\rangle$ allows to study the influence of cavity photons on the qubit. Furthermore, for known $g_{j}$, a measurement of $\tilde{\omega}_{\rm q}$ as a function of the probe tone power $P_{\rm rf}$ yields a calibration of the intracavity photon number $\bar{n}_{j}$ using the relation $P_{\rm rf} = \bar{n}_{j}\hbar\omega_{j}\kappa_{j}$, where $\kappa_{j}/2\pi$ is the full-width at half-maximum of the cavity resonance.

\section{Spectroscopy in the strong coupling limit\label{sec:spectroscopy}}

\begin{figure}[tb]
\includegraphics[width=0.9\linewidth]{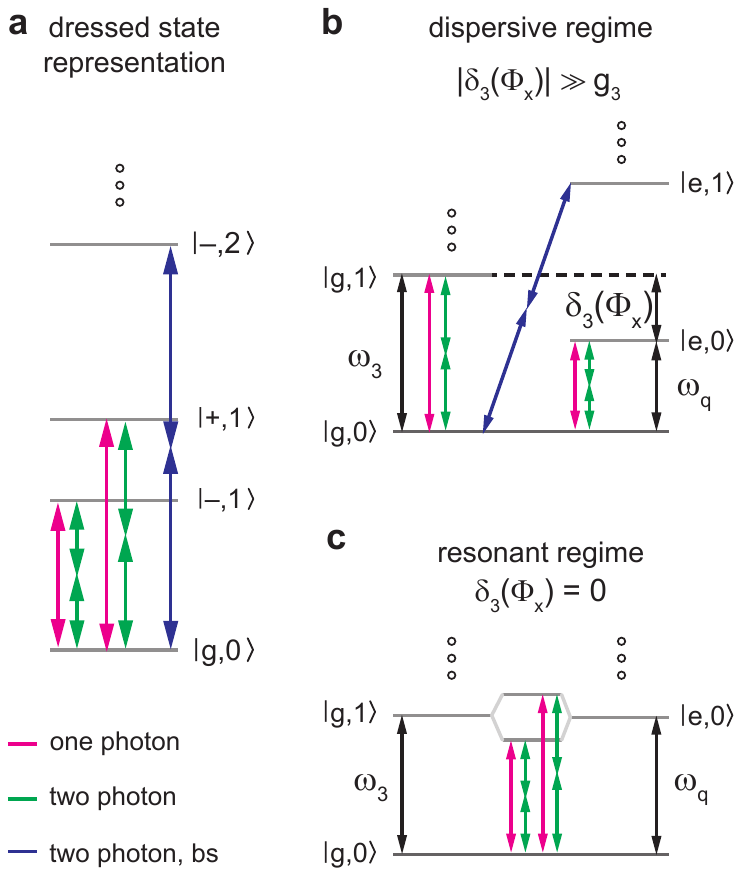}
\caption{(Color online) (a) Dressed state level diagram of the coupled qubit-cavity system. The states $|\pm,n\rangle$ denote the eigenstates of a single-mode Jaynes-Cummings
Hamiltonian with $n$ excitations. The transitions driven in the spectroscopy experiments neglecting selection rules are indicated by colored arrows: one-photon (magenta), two-photon (green) and two-photon blue sideband (bs, blue). A detailed discussion of the selection rules applicable to these
transitions is given in section~\ref{sec:symmetry}. (b) Dressed state level diagram in the dispersive limit ($|\delta_{\rm 3}| \gg g_{\rm 3}$), where $|-,1\rangle \rightarrow |g,1\rangle$ and $|+,1\rangle \rightarrow |e,0\rangle$. (c) Dressed
state level diagram for the resonant case ($\delta_{\rm 3} = 0)$. Here, the eigenstates $|\pm,1\rangle$ are symmetric (+) and antisymmetric (-) superpositions of $|g,1\rangle$ and $|e,0\rangle$.}\label{FIG3}
\end{figure}

\noindent Throughout this section we investigate our coupled qubit-cavity system using the $\lambda$-mode as a readout mode for microwave spectroscopy. Figure~\ref{FIG3}\,a shows the dressed state level diagram for the single-mode
Jaynes-Cummings Hamiltonian combining the qubit and the $3\lambda/2$-mode ($\omega_{\rm 3}/2\pi = \SI{7.775}{\giga\hertz}$). In Fig.~\ref{FIG3}\,b and c,
the dispersive and resonant limits are illustrated in more detail, respectively. The arrows in Fig.~\ref{FIG3} indicate the one- and two-photon transitions which we are driving in the 

\noindent spectroscopy experiments neglecting any selection rules. For the analysis presented in this section, we introduce the
notation $|\pm,n\rangle$ for the eigenstates of the single-mode Jaynes-Cummings Hamiltonian, where
\begin{align}
|-,n\rangle & =  \cos\Theta|g,n\rangle - \sin\Theta|e,n-1\rangle  \label{nminus}\quad,\\ \nonumber \\
|+,n\rangle & =  \sin\Theta|g,n\rangle + \cos\Theta|e,n-1\rangle\label{nplus} \quad.
\end{align}
Here, $|q,n\rangle$ denotes a state with $n$ photons in the $3\lambda/2$-mode and  the $q=\{g,e\}$ describes the qubit ground or excited states. The mixing angle $\Theta$ is given by \begin{equation}
\Theta = \frac{1}{2}\arctan(2g_{\rm j}\sqrt{n}/\delta_{\rm j})\quad.
\end{equation}
\subsection{One-photon spectroscopy and determination of coupling strengths\label{subsec:onephoton}}

\begin{figure}[b]
\includegraphics[width=0.9\linewidth]{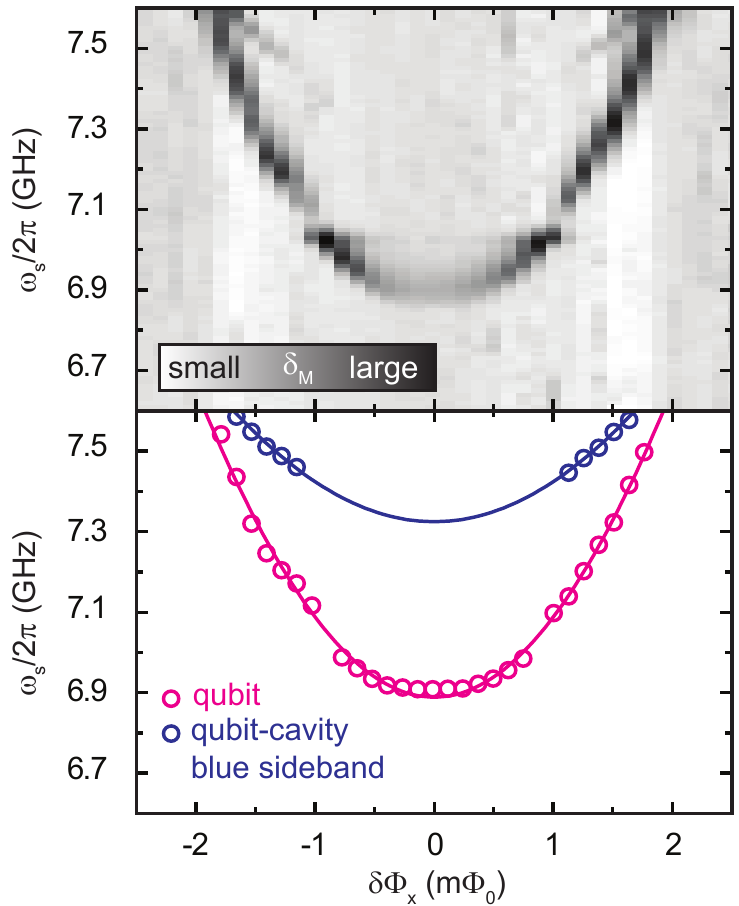}
\caption{(Color online) One-photon qubit spectroscopy showing the hyperbolic flux dependence of $\tilde{\omega}_{\rm q}$. In the top panel, the relative change in the transmission magnitude $\delta_{\rm M}$ (color-coded) is plotted as a function of  $\delta\Phi_{\rm x}$ and the spectroscopy frequency $\omega_{\rm s}/2\pi$. A blue sideband transition (resonance condition: $2\omega_{\rm s} = \omega_{\rm q} + \omega_{\rm 3}$) is visible. The magenta and blue circles in the bottom panel indicate the fitted center frequencies of the qubit spectroscopy signal and the blue sideband spectroscopy signal, respectively. The solid magenta line is a fit to the energy level spectrum of Hamiltonian (\ref{FullH}) while the solid blue line is evaluated using the fit parameters derived from the qubit data. The flux-independent feature in the region $\omega_{\rm s}/2\pi \approx \SI{7}{\giga\hertz}$ is attributed to the presence of a spurious fluctuator. The data presented in Fig.~\ref{FIG2} and Fig.~\ref{FIG4} is taken at an input power of $P_{\rm rf} = -134~\text{dBm}$ corresponding to
$\bar{n}_{\rm 2} < 0.94$.}\label{FIG4}
\end{figure}

\noindent The one-photon spectroscopy data is shown in Fig.~\ref{FIG4} and reflects the hyperbolic flux dependence of the qubit transition frequency. From the spectrum we extract the qubit parameters $\Delta/h = \SI{6.88}{\giga\hertz}$ and
$2I_{\rm p} = \SI{532}{\nano\ampere}$. Knowing $\tilde{\omega}_{\rm q}$, the qubit-resonator coupling rates $g_{j}$ can be determined by fitting the cavity transmission spectra (see Fig.~\ref{FIG2}\,b to the energy level spectrum of the Hamiltonian (\ref{FullH}).
\noindent The fitted coupling rates are $g_{\rm 2}/2\pi = \SI{106.5}{\mega\hertz}$ and $g_{\rm 3}/2\pi = \SI{90.7}{\mega\hertz}$. In the low power limit ($P_{\rm rf}$, $P_{\rm s} \rightarrow 0$), the full-width at
half maximum (FWHM) of the qubit spectroscopy signal is $\gamma \approx \SI{50}{\mega\hertz}$ and all decay rates $\kappa_{j}$ of the resonator modes are smaller than \SI{1.4}{\mega\hertz}. Thus, since $g_{j} > \kappa_{j}, \gamma$  the qubit-resonator system is in the strong coupling limit. In order to ensure a correct determination of $\Delta$ and $g_{j}$, the transmission spectra and the spectroscopy data shown in Fig.~\ref{FIG2}\,b and Fig.~\ref{FIG4}, respectively, were recorded in the low-power limit ($\bar{n}_{j} < 0.94$). The photon number was calibrated\cite{Schuster:2005a,Abdumalikov:2008} by measuring the photon number dependent ac-Zeeman shift of $\tilde{\omega}_{\rm q}$ as described in section~\ref{sec:setup}. Two additional features can be clearly identified in the one-photon spectroscopy data: first, around $\omega_{\rm s}/2\pi \approx \SI{7.00}{\giga\hertz}$, a flux-independent signature is visible and can be attributed to a spurious fluctuator present in our system. Such fluctuators can be either resonant modes or microscopic two-level systems and can significantly degrade the coherence properties of solid-state based artificial atoms.\cite{Simmonds:2004} Furthermore, it was shown that the presence of two-level fluctuators can lead to symmetry breaking,\cite{Deppe:2008a} however, this was not the case in our experiment and for the remainder of this paper we will therefore neglect the presence of the fluctuator. Second, in the region from $\SI{7.4}{\giga\hertz} \lesssim \omega_{\rm s}/2\pi \lesssim \SI{7.6}{\giga\hertz}$ our data reveals a flux-dependent spectroscopic signature which is consistent with a two-photon driven blue-sideband transition of the qubit and the $3\lambda/2$-mode. The selection rules for such transitions will be discussed in section~\ref{sec:symmetry}.

\begin{figure*}[tb]
\includegraphics[width=0.9\linewidth]{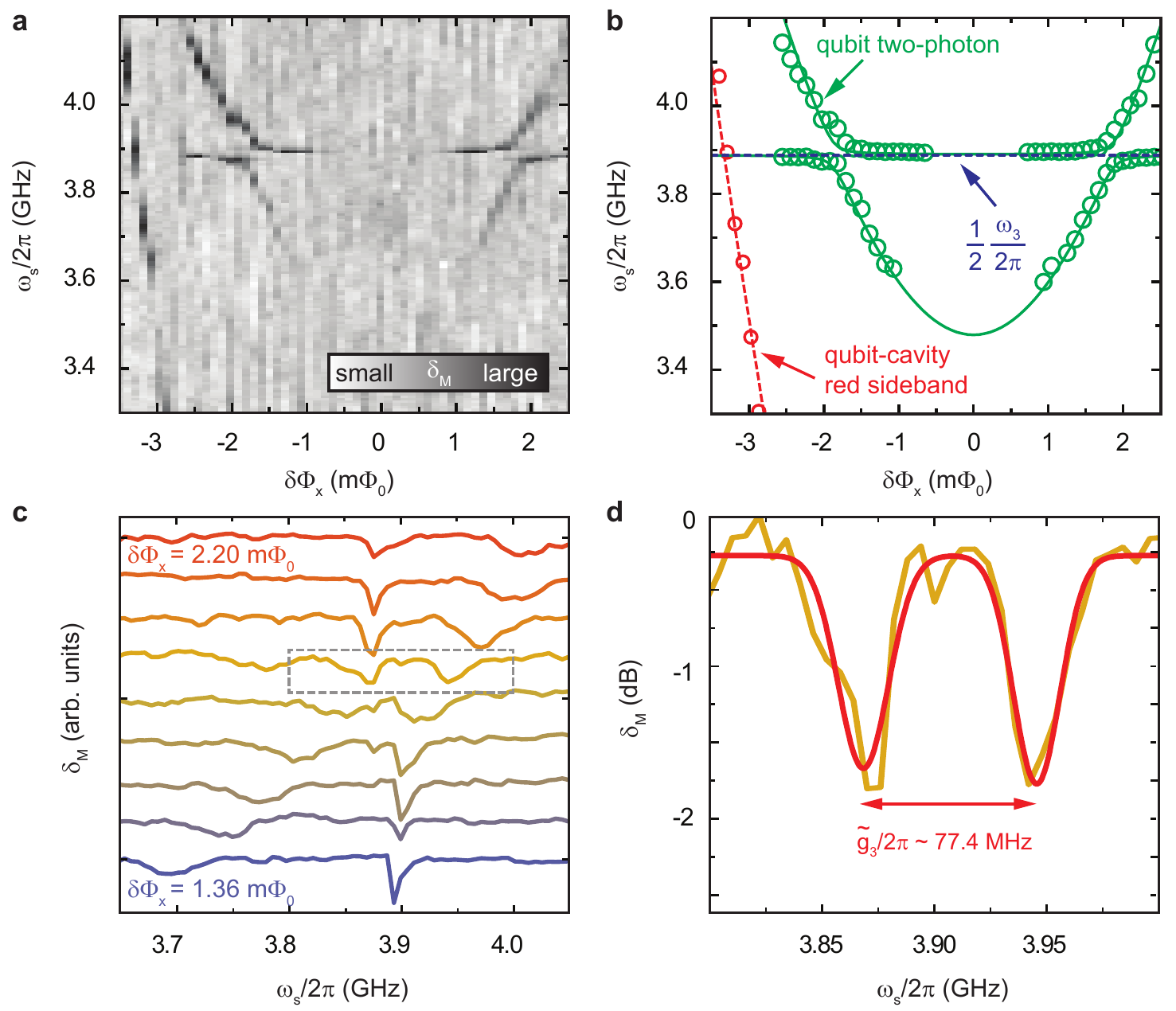}
\caption{(Color online) (a) Two-photon spectroscopy. The relative change in the transmission magnitude at the probe tone frequency $\omega_{\rm rf}$ is plotted as a function of $\delta\Phi_{\rm x}$ and the spectroscopy frequency $\omega_{\rm s}/2\pi$. At $\delta\Phi_{\rm x} = 0$, two-photon transitions are strictly
forbidden due to selection rules originating in the symmetry of the qubit potential. Away from the degeneracy point, the magnitude of the spectroscopy signal increases. The two anticrossings are the spectroscopic correspondents of
the qubit-cavity anticrossings visible in the cavity transmission spectrum of the $3\lambda/2$-mode (see Fig.~\ref{FIG2}\,b, top panel). On the left hand
side, a one-photon driven red sideband transition (resonance condition: $\omega_{\rm s} = \omega_{q} - \omega_{\rm 2}$) is visible. This transition corresponds to an exchange of an excitation between the qubit and the $\lambda$-mode. (b) Fitted center frequencies of the spectroscopic qubit response (green circles) as a function of $\delta\Phi_{\rm x}$. The solid green line is the numerically evaluated energy level spectrum of Hamiltonian (\ref{FullH}) with $\omega_{\rm q} \rightarrow \omega_{\rm q}/2, \omega_{\rm n} \rightarrow \omega_{\rm n}/2$ and $g_{n} \rightarrow g_{n}/2$. Dotted blue line: $1/2 \cdot \omega_{\rm 3}/2\pi$. Dashed red line: calculated flux
dependence of the red sideband transition. (c) Two-photon spectroscopy signal in the anticrossing region plotted as a function of $\omega_{\rm s}/2\pi$. The curves for different constant $\delta\Phi_{\rm x}$ are shifted along the ordinate. The gray dashed box marks the area magnified in (d). (d) Two-photon spectroscopy signal at a flux value where the qubit is resonant with the $3\lambda/2$-mode. The red solid line represents a fit to the sum of two Lorentzians from which we extract a dip separation of $\tilde{g}_{\rm 3} =
\SI{77.4}{\mega\hertz}$.}\label{FIG5}
\end{figure*}

\subsection{Two-photon spectroscopy\label{subsec:twophoton}}

\noindent We now investigate the spectroscopic response of the qubit-cavity system under direct two-photon driving. Again we use the flux dependent transmission maximum of the $\lambda$-mode as probe frequency. However, for a two-photon drive, the spectroscopy tone is applied in a frequency range around $2\omega_{\rm s} = \tilde{\omega}_{\rm q}$. The color-coded data and the fitted center frequencies of the spectroscopy transmission signal are shown in Fig.~\ref{FIG5}\,a and b, respectively. The spectrum shows a signature with a hyperbolic flux dependence, two anticrossings located symmetrically around $\delta\Phi_{\rm x} = 0$, and
additional flux-independent features in close vicinity of the anticrossing regions. A fit of the spectrum (see Fig.~\ref{FIG5}\,b) yields the two-photon qubit parameters $\Delta^{\rm 2ph}/h = \SI{3.48}{\giga\hertz}$ and $2I^{\rm
2ph}_{\rm p} = \SI{296}{\nano\ampere}$ which are in good agreement with the one-photon spectroscopy data. We attribute the deviations from the expected values ($\Delta^{\rm 2ph} = \Delta/2$ and $I^{\rm
2ph}_{\rm p} = I_{\rm p}/2$) to a higher intracavity photon number ($\bar{n}_{\rm 2} \approx 2.2$) compared to the one-photon spectroscopy measurement, thereby inducing an
ac-Zeeman shift in $\tilde{\omega}_{\rm q}$ towards higher frequencies.

\subsection{Symmetry properties and selection rules}\label{sec:symmetry}

\noindent The gradual disappearance of the qubit spectroscopy signal (see Fig.~\ref{FIG5}\,a) for $\delta\Phi_{\rm x} \rightarrow 0$ can be understood by considering the symmetry properties of our system which imply selection rules
for the allowed transitions. Selection rules are intimately related to the quantum mechanical concept of parity. The parity operator $\hat\Pi$ divides the set of all states into three groups: even states $(|\psi_{\rm +}\rangle)$, odd states $(|\psi_{\rm -}\rangle)$ and states without well-defined parity. The former two are eigenstates of $\hat\Pi$ with eigenvalues $+1$ and $-1$, respectively. In a similar fashion, it is possible to divide the set of all operators into similar classes. While an even operator $\hat{A}_{\rm +}$ commutes with $\hat\Pi$, an odd operator $\hat{A}_{\rm -}$ anticommutes with
$\hat\Pi$. It can be shown,~\cite{CohenTannoudji:2005} that the matrix elements of an even operator are zero between states of different parity
\begin{equation}
\langle\psi_{\rm +}|\hat{A}_{\rm +}|\psi_{\rm -}\rangle =
\langle\psi_{\rm -}|\hat{A}_{\rm +}|\psi_{\rm +}\rangle = 0
\end{equation}
\noindent while the matrix elements of an odd operator are zero between states of the same parity
\begin{equation}
\langle\psi_{\rm +}|\hat{A}_{\rm -}|\psi_{\rm +}\rangle =
\langle\psi_{\rm -}|\hat{A}_{\rm -}|\psi_{\rm -}\rangle = 0\,.
\end{equation}
\noindent At $\delta\Phi_{\rm x} = 0$, the two lowest qubit energy eigenstates $|g\rangle$ and $|e\rangle$ represent symmetric and antisymmetric superpositions of the persistent current states $|\pm I_{\rm p}\rangle$. Transitions between $|g\rangle$ and $|e\rangle$ are induced by an external
drive $\propto (\hat{a}^{\dagger} + \hat{a})$ with respect to a frame rotating at the drive frequency. It is easy to verify that the drive $(\hat{a}^{\dagger} + \hat{a})$ is an odd parity operator and as $|g\rangle$ and $|e\rangle$ correspond to states with different parities, one-photon transitions are
allowed at the qubit degeneracy point. On the other hand, a two-photon ($n$-photon) transition is equivalent to the application of two ($n$) subsequent drives with frequency $\tilde{\omega}_{\rm q}/2$ ($\tilde{\omega}_{\rm q}/n$). With
\begin{equation}
(\hat\Pi\hat{A}_{\rm -}\hat\Pi)^n = (-1)^n\hat{A}_{\rm -}^n\,,
\end{equation}
\noindent the two-photon drive is effectively an even parity operator and two-photon transitions between $|g\rangle$ and $|e\rangle$ are forbidden. Away from the qubit degeneracy point ($\delta\Phi_{\rm x} \neq 0$), the parity of the Hamiltonian (\ref{FullH}) is not well-defined. In the dispersive limit, the effective Hamiltonian can be derived\cite{Deppe:2008a,Deppe:2009PhD} by means
of a Schrieffer-Wolff transformation yielding finite transition matrix elements for one photon
\begin{equation}
\frac{\Omega}{4}\sin\theta
\end{equation}
\noindent and two photon drive
\begin{equation}
\frac{\Omega^2}{4\Delta}\sin^2\theta\cos\theta\quad.
\end{equation}
\noindent When deriving the above expressions, we assume the drive Hamiltonian $\hat{H}_{\rm d} = (\Omega/2)\hat{\sigma}_{z}\cos(\omega t)$. Here, $\omega = \omega_{\rm q}/n$ for an $n$-photon drive and $\Omega$ refers to the driving strength. The $\cos\theta$-dependence of the two-photon spectroscopy signal as qualitatively reproduced by our data (see Fig.~\ref{FIG5}\,a) explains the gradual disappearance of the qubit signature as $\cos\theta \rightarrow 0$ for
$\delta\Phi_{\rm x} \rightarrow 0$. The selection rules for one- and two-photon driven red and blue sideband transitions can be derived in a similar fashion.~\cite{Blais:2007} The two-photon blue sideband illustrated in Fig.~\ref{FIG4}
corresponds to the transition $|g,0\rangle \rightarrow |e,1\rangle$. These states have the same parity and since the two-photon drive has even parity, the transition is allowed for arbitrary $\delta\Phi_{\rm x}$. However, because this
is a second-order transition, its amplitude is small and the spectroscopic signature can not be resolved within our measurement resolution and for low spectroscopy power $P_{\rm s}$. Finally we note that for the harmonic resonator potential multi-photon transitions are forbidden,~\cite{Deppe:2008a}
which is illustrated by the absence of a flux-independent spectroscopic feature at $\omega_{\rm s} = \omega_{\rm 3}/2$ except for the anticrossing regions. Here, a flux-independent resonator-like signature is visible due to the large
qubit-like component of the eigenstates of the Jaynes-Cummings Hamiltonian (see Eqs.~(\ref{nminus}) and (\ref{nplus})).

\subsection{Anticrossing under two-photon driving}

\noindent The anticrossings in the spectrum shown in Fig.~\ref{FIG5}\,a are a direct manifestation of the one-photon qubit-resonator anticrossing (see Fig.~\ref{FIG2}\,a) under two-photon driving. Here, the qubit and resonator energies are degenerate, and the eigenstates $|\pm,1\rangle$ of the coupled system are symmetric and antisymmetric superpositions of $|g,1\rangle$ and
$|e,0\rangle$ (see Fig.~\ref{FIG3}\,c). The fact that these anticrossings are resolved so clearly indicates that the qubit-cavity system is in the strong coupling limit. There are several ways to determine $g_{j}$, all requiring a negligible photon population of the cavity modes. A straightforward way to investigate the coupling is to record the vacuum Rabi splitting. Observing this splitting is experimentally challenging but has been demonstrated with different kinds of
qubits.~\cite{Wallraff:2004a,Houck:2007a,Sillanpaa:2007a,Fedorov:2010,Steffen:2010}
Typically, the amplitude of the two peaks corresponding to the entangled qubit-cavity states is significantly reduced compared to the bare cavity mode due to the following reason: on resonance, the two systems can exchange energy and a decay of the qubit into non-radiative channels eliminates the shared excitation of the joint system. Such processes thus lead to a significant reduction of the transmitted microwave amplitude which in turn increases the required signal averaging times. Furthermore, the Rabi peaks become inhomogeneously broadened with increasing power. The coupling can also be extracted from a fit of the energy level spectrum of the qubit-cavity
Hamiltonian (see section~\ref{subsec:onephoton}) to the measured cavity transmission spectra. Here, we extract the qubit-cavity coupling from the dispersive two-photon spectroscopy data. This has the advantage that both drive
frequencies, $\omega_{\rm rf}$ and $\omega_{\rm s}$ are applied off-resonant to the cavity mode resonant with the qubit (in our case the $3\lambda/2$-mode). Therefore, this mode is not populated significantly by the drives and the vacuum Rabi splitting can be determined form the spectroscopy data.
Figure~\ref{FIG5}\,c shows the measured relative change of transmission magnitude as a function of $\delta\Phi_{\rm x}$ and $\omega_{\rm s}$ for the anticrossing located at $\delta\Phi_{\rm x} \approx 1.9~\text{m}\Phi_{\rm 0}$. For
increasing flux detuning, $\tilde{\omega}_{\rm q}$ is shifted towards higher frequencies until the resonance condition $\tilde{\omega}_{\rm q} = \omega_{\rm 3}$ is fulfilled. We extract a minimum separation of the two dips of $\tilde{g}_{\rm 3}/2\pi = \SI{77.4}{\mega\hertz}$ as shown in Fig.~\ref{FIG5}\,d. Taking into account that $\sin\theta \approx 0.89$ at this flux value, we obtain $g_{\rm 3}/2\pi \approx \SI{87.0}{\mega\hertz}$ in good agreement with the fitted coupling rate determined from the cavity transmission
and the one-photon qubit spectroscopy experiments. From the FWHM $\approx (\gamma + \kappa_{\rm 3})/2$ of the spectroscopy signal we can estimate $\gamma/2\pi \approx \SI{43}{\mega\hertz}$ in agreement with the linewidth obtained from low-power one-photon spectroscopy.

\section{Conclusions\label{sec:conclusions}}

\noindent In summary, we investigate experimentally a superconducting flux qubit strongly coupled to a microwave transmission line resonator under one- and two-photon driving. At the qubit degeneracy point, the selection rules for dipole transitions are equivalent to the selection rules for transitions in natural atoms. By tuning the external control parameter $\delta\Phi_{\rm x}$ the symmetry of the system can be broken in a controlled way and two-photon driving of the coupled
qubit-cavity system becomes possible. This technique allows to extract the coupling rates $g_{\rm n}$ of modes resonant with the qubit in the dispersive limit.\\

\begin{acknowledgments}
\noindent This work is supported by the German Research Foundation through SFB~631, the German Excellence Initiative via NIM and by the EU through the Marie Curie Initial Training Network CCQED. J\,.J\,.G\,.R\,. acknowledges financial support from Spanish MICINN Project FIS2009-10061 and CAM research consortium QUITEMAD Grant No.\,S2009-ESP-1594. E\,.S\,. acknowledges financial support from Basque Government grant IT472-10, MICINN FIS2009-12773-C02-01, SOLID and CCQED European projects.
\end{acknowledgments}


\begin{thebibliography}{10}%
\makeatletter
\providecommand \@ifxundefined [1]{%
 \ifx #1\undefined \expandafter \@firstoftwo
 \else \expandafter \@secondoftwo
\fi
}%
\providecommand \@ifnum [1]{%
 \ifnum #1\expandafter \@firstoftwo
 \else \expandafter \@secondoftwo
\fi
}%
\providecommand \enquote [1]{``#1''}%
\providecommand \bibnamefont  [1]{#1}%
\providecommand \bibfnamefont [1]{#1}%
\providecommand \citenamefont [1]{#1}%
\providecommand\href[0]{\@sanitize\@href}%
\providecommand\@href[1]{\endgroup\@@startlink{#1}\endgroup\@@href}%
\providecommand\@@href[1]{#1\@@endlink}%
\providecommand \@sanitize [0]{\begingroup\catcode`\&12\catcode`\#12\relax}%
\@ifxundefined \pdfoutput {\@firstoftwo}{%
 \@ifnum{\z@=\pdfoutput}{\@firstoftwo}{\@secondoftwo}%
}{%
 \providecommand\@@startlink[1]{\leavevmode}%
 \providecommand\@@endlink[0]{}%
}{%
 \providecommand\@@startlink[1]{%
  \leavevmode
  \pdfstartlink
   attr{/Border[0 0 1 ]/H/I/C[0 1 1]}%
   user{/Subtype/Link/A<</Type/Action/S/URI/URI(#1)>>}%
  \relax
 }%
 \providecommand\@@endlink[0]{\pdfendlink}%
}%
\providecommand \url  [0]{\begingroup\@sanitize \@url }%
\providecommand \@url [1]{\endgroup\@href {#1}{\urlprefix}}%
\providecommand \urlprefix [0]{URL }%
\providecommand \Eprint[0]{\href }%
\@ifxundefined \urlstyle {%
  \providecommand \doi [1]{doi:\discretionary{}{}{}#1}%
}{%
  \providecommand \doi [0]{doi:\discretionary{}{}{}\begingroup
  \urlstyle{rm}\Url }%
}%
\providecommand \doibase [0]{http://dx.doi.org/}%
\providecommand \Doi[1]{\href{\doibase#1}}%
\providecommand \bibAnnote [3]{%
  \BibitemShut{#1}%
  \begin{quotation}\noindent
    \textsc{Key:}\ #2\\\textsc{Annotation:}\ #3%
  \end{quotation}%
}%
\providecommand \bibAnnoteFile [2]{%
  \IfFileExists{#2}{\bibAnnote {#1} {#2} {\input{#2}}}{}%
}%
\providecommand \typeout [0]{\immediate \write \m@ne }%
\providecommand \selectlanguage [0]{\@gobble}%
\providecommand \bibinfo [0]{\@secondoftwo}%
\providecommand \bibfield [0]{\@secondoftwo}%
\providecommand \translation [1]{[#1]}%
\providecommand \BibitemOpen[0]{}%
\providecommand \bibitemStop [0]{}%
\providecommand \bibitemNoStop [0]{.\EOS\space}%
\providecommand \EOS [0]{\spacefactor3000\relax}%
\providecommand \BibitemShut [1]{\csname bibitem#1\endcsname}%
\bibitem{Wallraff:2004a}%
  \BibitemOpen
  \bibfield{author}{%
  \bibinfo {author} {\bibfnamefont{A.}~\bibnamefont{Wallraff}}, \bibinfo
  {author} {\bibfnamefont{D.~I.}\ \bibnamefont{Schuster}}, \bibinfo {author}
  {\bibfnamefont{A.}~\bibnamefont{Blais}}, \bibinfo {author}
  {\bibfnamefont{L.}~\bibnamefont{Frunzio}}, \bibinfo {author}
  {\bibfnamefont{R.-S.}\ \bibnamefont{Huang}}, \bibinfo {author}
  {\bibfnamefont{J.}~\bibnamefont{Majer}}, \bibinfo {author}
  {\bibfnamefont{S.}~\bibnamefont{Kumar}}, \bibinfo {author}
  {\bibfnamefont{S.~M.}\ \bibnamefont{Girvin}},\ and\ \bibinfo {author}
  {\bibfnamefont{R.~J.}\ \bibnamefont{Schoelkopf}},\ }%
  \bibfield{journal}{%
  \bibinfo {journal} {Nature}\ }%
  \textbf{\bibinfo {volume} {431}},\ \bibinfo {pages} {162} (\bibinfo {year}
  {2004})%
  \bibAnnoteFile{NoStop}{Wallraff:2004a}%
\bibitem{Schuster:2007a}%
  \BibitemOpen
  \bibfield{author}{%
  \bibinfo {author} {\bibfnamefont{D.~I.}\ \bibnamefont{Schuster}}, \bibinfo
  {author} {\bibfnamefont{A.~A.}\ \bibnamefont{Houck}}, \bibinfo {author}
  {\bibfnamefont{J.~A.}\ \bibnamefont{Schreier}}, \bibinfo {author}
  {\bibfnamefont{A.}~\bibnamefont{Wallraff}}, \bibinfo {author}
  {\bibfnamefont{J.~M.}\ \bibnamefont{Gambetta}}, \bibinfo {author}
  {\bibfnamefont{A.}~\bibnamefont{Blais}}, \bibinfo {author}
  {\bibfnamefont{L.}~\bibnamefont{Frunzio}}, \bibinfo {author}
  {\bibfnamefont{J.}~\bibnamefont{Majer}}, \bibinfo {author}
  {\bibfnamefont{B.}~\bibnamefont{Johnson}}, \bibinfo {author}
  {\bibfnamefont{M.~H.}\ \bibnamefont{Devoret}}, \bibinfo {author}
  {\bibfnamefont{S.~M.}\ \bibnamefont{Girvin}},\ and\ \bibinfo {author}
  {\bibfnamefont{R.~J.}\ \bibnamefont{Schoelkopf}},\ }%
  \bibfield{journal}{%
  \bibinfo {journal} {Nature}\ }%
  \textbf{\bibinfo {volume} {445}},\ \bibinfo {pages} {515} (\bibinfo {year}
  {2007})%
  \bibAnnoteFile{NoStop}{Schuster:2007a}%
\bibitem{Houck:2007a}%
  \BibitemOpen
  \bibfield{author}{%
  \bibinfo {author} {\bibfnamefont{A.~A.}\ \bibnamefont{Houck}}, \bibinfo
  {author} {\bibfnamefont{D.~I.}\ \bibnamefont{Schuster}}, \bibinfo {author}
  {\bibfnamefont{J.~M.}\ \bibnamefont{Gambetta}}, \bibinfo {author}
  {\bibfnamefont{J.~A.}\ \bibnamefont{Schreier}}, \bibinfo {author}
  {\bibfnamefont{B.~R.}\ \bibnamefont{Johnson}}, \bibinfo {author}
  {\bibfnamefont{J.~M.}\ \bibnamefont{Chow}}, \bibinfo {author}
  {\bibfnamefont{L.}~\bibnamefont{Frunzio}}, \bibinfo {author}
  {\bibfnamefont{J.}~\bibnamefont{Majer}}, \bibinfo {author}
  {\bibfnamefont{M.~H.}\ \bibnamefont{Devoret}}, \bibinfo {author}
  {\bibfnamefont{S.~M.}\ \bibnamefont{Girvin}},\ and\ \bibinfo {author}
  {\bibfnamefont{R.~J.}\ \bibnamefont{Schoelkopf}},\ }%
  \bibfield{journal}{%
  \bibinfo {journal} {Nature}\ }%
  \textbf{\bibinfo {volume} {449}},\ \bibinfo {pages} {328} (\bibinfo {year}
  {2007})%
  \bibAnnoteFile{NoStop}{Houck:2007a}%
\bibitem{Astafiev:2007a}%
  \BibitemOpen
  \bibfield{author}{%
  \bibinfo {author} {\bibfnamefont{O.}~\bibnamefont{Astafiev}}, \bibinfo
  {author} {\bibfnamefont{K.}~\bibnamefont{Inomata}}, \bibinfo {author}
  {\bibfnamefont{A.~O.}\ \bibnamefont{Niskanen}}, \bibinfo {author}
  {\bibfnamefont{T.}~\bibnamefont{Yamamoto}}, \bibinfo {author}
  {\bibfnamefont{{\relax Yu}.~A.}\ \bibnamefont{Pashkin}}, \bibinfo {author}
  {\bibfnamefont{Y.}~\bibnamefont{Nakamura}},\ and\ \bibinfo {author}
  {\bibfnamefont{J.~S.}\ \bibnamefont{Tsai}},\ }%
  \bibfield{journal}{%
  \bibinfo {journal} {Nature}\ }%
  \textbf{\bibinfo {volume} {449}},\ \bibinfo {pages} {588} (\bibinfo {year}
  {2007})%
  \bibAnnoteFile{NoStop}{Astafiev:2007a}%
\bibitem{Abdumalikov:2008}%
  \BibitemOpen
  \bibfield{author}{%
  \bibinfo {author} {\bibfnamefont{A.}~\bibnamefont{Abdumalikov}}, \bibinfo
  {author} {\bibfnamefont{O.}~\bibnamefont{Astafiev}}, \bibinfo {author}
  {\bibfnamefont{Y.}~\bibnamefont{Nakamura}}, \bibinfo {author}
  {\bibfnamefont{Y.}~\bibnamefont{Pashkin}},\ and\ \bibinfo {author}
  {\bibfnamefont{J.}~\bibnamefont{Tsai}},\ }%
  \bibfield{journal}{%
  \bibinfo {journal} {Phys.\ Rev.\ B}\ }%
  \textbf{\bibinfo {volume} {78}},\ \bibinfo {pages} {180502} (\bibinfo {year}
  {2008})%
  \bibAnnoteFile{NoStop}{Abdumalikov:2008}%
\bibitem{Hofheinz:2008a}%
  \BibitemOpen
  \bibfield{author}{%
  \bibinfo {author} {\bibfnamefont{M.}~\bibnamefont{Hofheinz}}, \bibinfo
  {author} {\bibfnamefont{E.~M.}\ \bibnamefont{Weig}}, \bibinfo {author}
  {\bibfnamefont{M.}~\bibnamefont{Ansmann}}, \bibinfo {author}
  {\bibfnamefont{R.~C.}\ \bibnamefont{Bialczak}}, \bibinfo {author}
  {\bibfnamefont{E.}~\bibnamefont{Lucero}}, \bibinfo {author}
  {\bibfnamefont{M.}~\bibnamefont{Neeley}}, \bibinfo {author}
  {\bibfnamefont{A.~D.}\ \bibnamefont{O'Connell}}, \bibinfo {author}
  {\bibfnamefont{H.}~\bibnamefont{Wang}}, \bibinfo {author}
  {\bibfnamefont{J.~M.}\ \bibnamefont{Martinis}},\ and\ \bibinfo {author}
  {\bibfnamefont{A.~N.}\ \bibnamefont{Cleland}},\ }%
  \bibfield{journal}{%
  \bibinfo {journal} {Nature}\ }%
  \textbf{\bibinfo {volume} {454}},\ \bibinfo {pages} {310} (\bibinfo {year}
  {2008})%
  \bibAnnoteFile{NoStop}{Hofheinz:2008a}%
\bibitem{Niemczyk:2010}%
  \BibitemOpen
  \bibfield{author}{%
  \bibinfo {author} {\bibfnamefont{T.}~\bibnamefont{Niemczyk}}, \bibinfo
  {author} {\bibfnamefont{F.}~\bibnamefont{Deppe}}, \bibinfo {author}
  {\bibfnamefont{H.}~\bibnamefont{Huebl}}, \bibinfo {author}
  {\bibfnamefont{E.~P.}\ \bibnamefont{Menzel}}, \bibinfo {author}
  {\bibfnamefont{F.}~\bibnamefont{Hocke}}, \bibinfo {author}
  {\bibfnamefont{M.~J.}\ \bibnamefont{Schwarz}}, \bibinfo {author}
  {\bibfnamefont{J.~J.}\ \bibnamefont{Garcia-Ripoll}}, \bibinfo {author}
  {\bibfnamefont{D.}~\bibnamefont{Zueco}}, \bibinfo {author}
  {\bibfnamefont{T.}~\bibnamefont{H\"{u}mmer}}, \bibinfo {author}
  {\bibfnamefont{E.}~\bibnamefont{Solano}}, \bibinfo {author}
  {\bibfnamefont{A.}~\bibnamefont{Marx}},\ and\ \bibinfo {author}
  {\bibfnamefont{R.}~\bibnamefont{Gross}},\ }%
  \bibfield{journal}{%
  \bibinfo {journal} {Nature Physics}\ }%
  \textbf{\bibinfo {volume} {6}},\ \bibinfo {pages} {772} (\bibinfo {year}
  {2010})%
  \bibAnnoteFile{NoStop}{Niemczyk:2010}%
\bibitem{Haroche:2006a}%
  \BibitemOpen
  \bibfield{author}{%
  \bibinfo {author} {\bibfnamefont{S.}~\bibnamefont{Haroche}}\ and\ \bibinfo
  {author} {\bibfnamefont{J.-M.}\ \bibnamefont{Raimond}},\ }%
  \emph{\bibinfo {title} {Exploring the Quantum}}\ (\bibinfo {publisher}
  {Oxford University Press Inc.},\ \bibinfo {address} {New York},\ \bibinfo
  {year} {2006})%
  \bibAnnoteFile{NoStop}{Haroche:2006a}%
\bibitem{Ilichev:2010a}%
  \BibitemOpen
  \bibfield{author}{%
  \bibinfo {author} {\bibfnamefont{E.}~\bibnamefont{Il'ichev}}, \bibinfo
  {author} {\bibfnamefont{S.~N.}\ \bibnamefont{Shevchenko}}, \bibinfo {author}
  {\bibfnamefont{S.~H.~W.}\ \bibnamefont{van~der Ploeg}}, \bibinfo {author}
  {\bibfnamefont{M.}~\bibnamefont{Grajcar}}, \bibinfo {author}
  {\bibfnamefont{E.~A.}\ \bibnamefont{Temchenko}}, \bibinfo {author}
  {\bibfnamefont{A.~N.}\ \bibnamefont{Omelyanchouk}},\ and\ \bibinfo {author}
  {\bibfnamefont{H.-G.}\ \bibnamefont{Meyer}},\ }%
  \bibfield{journal}{%
  \bibinfo {journal} {Phys. Rev. B}\ }%
  \textbf{\bibinfo {volume} {81}},\ \bibinfo {pages} {012506} (\bibinfo {year}
  {2010})%
  \bibAnnoteFile{NoStop}{Ilichev:2010a}%
\bibitem{Oliver:2005a}%
  \BibitemOpen
  \bibfield{author}{%
  \bibinfo {author} {\bibfnamefont{W.~D.}\ \bibnamefont{Oliver}}, \bibinfo
  {author} {\bibfnamefont{Y.}~\bibnamefont{Yu}}, \bibinfo {author}
  {\bibfnamefont{J.~C.}\ \bibnamefont{Lee}}, \bibinfo {author}
  {\bibfnamefont{K.~K.}\ \bibnamefont{Berggren}}, \bibinfo {author}
  {\bibfnamefont{L.~S.}\ \bibnamefont{Levitov}},\ and\ \bibinfo {author}
  {\bibfnamefont{T.~P.}\ \bibnamefont{Orlando}},\ }%
  \bibfield{journal}{%
  \bibinfo {journal} {Science}\ }%
  \textbf{\bibinfo {volume} {310}},\ \bibinfo {pages} {1653} (\bibinfo {year}
  {2005})%
  \bibAnnoteFile{NoStop}{Oliver:2005a}%
\bibitem{Sillanpaa:2006a}%
  \BibitemOpen
  \bibfield{author}{%
  \bibinfo {author} {\bibfnamefont{M.}~\bibnamefont{Sillanp{\"a}{\"a}}},
  \bibinfo {author} {\bibfnamefont{T.}~\bibnamefont{Lehtinen}}, \bibinfo
  {author} {\bibfnamefont{A.}~\bibnamefont{Paila}}, \bibinfo {author}
  {\bibfnamefont{Y.}~\bibnamefont{Makhlin}},\ and\ \bibinfo {author}
  {\bibfnamefont{P.}~\bibnamefont{Hakonen}},\ }%
  \bibfield{journal}{%
  \bibinfo {journal} {Phys. Rev. Lett.}\ }%
  \textbf{\bibinfo {volume} {96}},\ \bibinfo {pages} {187002} (\bibinfo {year}
  {2006})%
  \bibAnnoteFile{NoStop}{Sillanpaa:2006a}%
\bibitem{Lupascu:2009a}%
  \BibitemOpen
  \bibfield{author}{%
  \bibinfo {author} {\bibfnamefont{A.}~\bibnamefont{Lupa{\c{s}}cu}}, \bibinfo
  {author} {\bibfnamefont{P.}~\bibnamefont{Bertet}}, \bibinfo {author}
  {\bibfnamefont{E.~F.~C.}\ \bibnamefont{Driessen}}, \bibinfo {author}
  {\bibfnamefont{C.~J. P.~M.}\ \bibnamefont{Harmans}},\ and\ \bibinfo {author}
  {\bibfnamefont{J.~E.}\ \bibnamefont{Mooij}},\ }%
  \bibfield{journal}{%
  \bibinfo {journal} {Phys. Rev. B}\ }%
  \textbf{\bibinfo {volume} {80}},\ \bibinfo {pages} {172506} (\bibinfo {year}
  {2009})%
  \bibAnnoteFile{NoStop}{Lupascu:2009a}%
\bibitem{Berns:2005}%
  \BibitemOpen
  \bibfield{author}{%
  \bibinfo {author} {\bibfnamefont{D.~M.}\ \bibnamefont{Berns}}, \bibinfo
  {author} {\bibfnamefont{M.~S.}\ \bibnamefont{Rudner}}, \bibinfo {author}
  {\bibfnamefont{S.~O.}\ \bibnamefont{Valenzuela}}, \bibinfo {author}
  {\bibfnamefont{K.~K.}\ \bibnamefont{Berggren}}, \bibinfo {author}
  {\bibfnamefont{W.~D.}\ \bibnamefont{Oliver}}, \bibinfo {author}
  {\bibfnamefont{L.~S.}\ \bibnamefont{Levitov}},\ and\ \bibinfo {author}
  {\bibfnamefont{T.~P.}\ \bibnamefont{Orlando}},\ }%
  \bibfield{journal}{%
  \bibinfo {journal} {Nature}\ }%
  \textbf{\bibinfo {volume} {455}},\ \bibinfo {pages} {51} (\bibinfo {year}
  {2008})%
  \bibAnnoteFile{NoStop}{Berns:2005}%
\bibitem{Bushev:2010}%
  \BibitemOpen
  \bibfield{author}{%
  \bibinfo {author} {\bibfnamefont{P.}~\bibnamefont{Bushev}}, \bibinfo {author}
  {\bibfnamefont{C.}~\bibnamefont{M\"uller}}, \bibinfo {author}
  {\bibfnamefont{J.}~\bibnamefont{Lisenfeld}}, \bibinfo {author}
  {\bibfnamefont{J.~H.}\ \bibnamefont{Cole}}, \bibinfo {author}
  {\bibfnamefont{A.}~\bibnamefont{Lukashenko}}, \bibinfo {author}
  {\bibfnamefont{A.}~\bibnamefont{Shnirman}},\ and\ \bibinfo {author}
  {\bibfnamefont{A.~V.}\ \bibnamefont{Ustinov}},\ }%
  \bibfield{journal}{%
  \bibinfo {journal} {Phys. Rev. B}\ }%
  \textbf{\bibinfo {volume} {82}},\ \bibinfo {pages} {134530} (\bibinfo {year}
  {2010})%
  \bibAnnoteFile{NoStop}{Bushev:2010}%
\bibitem{Wallraff:2007a}%
  \BibitemOpen
  \bibfield{author}{%
  \bibinfo {author} {\bibfnamefont{A.}~\bibnamefont{Wallraff}}, \bibinfo
  {author} {\bibfnamefont{D.~I.}\ \bibnamefont{Schuster}}, \bibinfo {author}
  {\bibfnamefont{A.}~\bibnamefont{Blais}}, \bibinfo {author}
  {\bibfnamefont{J.~M.}\ \bibnamefont{Gambetta}}, \bibinfo {author}
  {\bibfnamefont{J.}~\bibnamefont{Schreier}}, \bibinfo {author}
  {\bibfnamefont{L.}~\bibnamefont{Frunzio}}, \bibinfo {author}
  {\bibfnamefont{M.~H.}\ \bibnamefont{Devoret}}, \bibinfo {author}
  {\bibfnamefont{S.~M.}\ \bibnamefont{Girvin}}, ,\ and\ \bibinfo {author}
  {\bibfnamefont{R.~J.}\ \bibnamefont{Schoelkopf}},\ }%
  \bibfield{journal}{%
  \bibinfo {journal} {Phys.\ Rev.\ Lett.}\ }%
  \textbf{\bibinfo {volume} {99}},\ \bibinfo {pages} {050501} (\bibinfo {year}
  {2007})%
  \bibAnnoteFile{NoStop}{Wallraff:2007a}%
\bibitem{Deppe:2008a}%
  \BibitemOpen
  \bibfield{author}{%
  \bibinfo {author} {\bibfnamefont{F.}~\bibnamefont{Deppe}}, \bibinfo {author}
  {\bibfnamefont{M.}~\bibnamefont{Mariantoni}}, \bibinfo {author}
  {\bibfnamefont{E.~P.}\ \bibnamefont{Menzel}}, \bibinfo {author}
  {\bibfnamefont{A.}~\bibnamefont{Marx}}, \bibinfo {author}
  {\bibfnamefont{S.}~\bibnamefont{Saito}}, \bibinfo {author}
  {\bibfnamefont{K.}~\bibnamefont{Kakuyanagi}}, \bibinfo {author}
  {\bibfnamefont{T.}~\bibnamefont{Meno}}, \bibinfo {author}
  {\bibfnamefont{K.}~\bibnamefont{Semba}}, \bibinfo {author}
  {\bibfnamefont{H.}~\bibnamefont{Takayanagi}}, \bibinfo {author}
  {\bibfnamefont{E.}~\bibnamefont{Solano}},\ and\ \bibinfo {author}
  {\bibfnamefont{R.}~\bibnamefont{Gross}},\ }%
  \bibfield{journal}{%
  \bibinfo {journal} {Nature Physics}\ }%
  \textbf{\bibinfo {volume} {4}},\ \bibinfo {pages} {686} (\bibinfo {year}
  {2008})%
  \bibAnnoteFile{NoStop}{Deppe:2008a}%
\bibitem{Niemczyk:2009a}%
  \BibitemOpen
  \bibfield{author}{%
  \bibinfo {author} {\bibfnamefont{T.}~\bibnamefont{Niemczyk}}, \bibinfo
  {author} {\bibfnamefont{F.}~\bibnamefont{Deppe}}, \bibinfo {author}
  {\bibfnamefont{M.}~\bibnamefont{Mariantoni}}, \bibinfo {author}
  {\bibfnamefont{E.~P.}\ \bibnamefont{Menzel}}, \bibinfo {author}
  {\bibfnamefont{E.}~\bibnamefont{Hoffmann}}, \bibinfo {author}
  {\bibfnamefont{G.}~\bibnamefont{Wild}}, \bibinfo {author}
  {\bibfnamefont{L.}~\bibnamefont{Eggenstein}}, \bibinfo {author}
  {\bibfnamefont{A.}~\bibnamefont{Marx}},\ and\ \bibinfo {author}
  {\bibfnamefont{R.}~\bibnamefont{Gross}},\ }%
  \bibfield{journal}{%
  \bibinfo {journal} {Supercond.\ Sci.\ Technol.}\ }%
  \textbf{\bibinfo {volume} {22}},\ \bibinfo {pages} {034009} (\bibinfo {year}
  {2009})%
  \bibAnnoteFile{NoStop}{Niemczyk:2009a}%
\bibitem{Liu:2005}%
  \BibitemOpen
  \bibfield{author}{%
  \bibinfo {author} {\bibfnamefont{Y.-x.}\ \bibnamefont{Liu}}, \bibinfo
  {author} {\bibfnamefont{J.~Q.}\ \bibnamefont{You}}, \bibinfo {author}
  {\bibfnamefont{L.~F.}\ \bibnamefont{Wei}}, \bibinfo {author}
  {\bibfnamefont{C.~P.}\ \bibnamefont{Sun}},\ and\ \bibinfo {author}
  {\bibfnamefont{F.}~\bibnamefont{Nori}},\ }%
  \bibfield{journal}{%
  \bibinfo {journal} {Phys. Rev. Lett.}\ }%
  \textbf{\bibinfo {volume} {95}},\ \bibinfo {pages} {087001} (\bibinfo {year}
  {2005})%
  \bibAnnoteFile{NoStop}{Liu:2005}%
\bibitem{Mooij:1999a}%
  \BibitemOpen
  \bibfield{author}{%
  \bibinfo {author} {\bibfnamefont{J.~E.}\ \bibnamefont{Mooij}}, \bibinfo
  {author} {\bibfnamefont{T.~P.}\ \bibnamefont{Orlando}}, \bibinfo {author}
  {\bibfnamefont{L.}~\bibnamefont{Levitov}}, \bibinfo {author}
  {\bibfnamefont{L.}~\bibnamefont{Tian}}, \bibinfo {author}
  {\bibfnamefont{C.~H.}\ \bibnamefont{van~der Wal}},\ and\ \bibinfo {author}
  {\bibfnamefont{S.}~\bibnamefont{Lloyd}},\ }%
  \bibfield{journal}{%
  \bibinfo {journal} {Science}\ }%
  \textbf{\bibinfo {volume} {285}},\ \bibinfo {pages} {1036} (\bibinfo {year}
  {1999})%
  \bibAnnoteFile{NoStop}{Mooij:1999a}%
\bibitem{Deppe:2007a}%
  \BibitemOpen
  \bibfield{author}{%
  \bibinfo {author} {\bibfnamefont{F.}~\bibnamefont{Deppe}}, \bibinfo {author}
  {\bibfnamefont{M.}~\bibnamefont{Mariantoni}}, \bibinfo {author}
  {\bibfnamefont{E.~P.}\ \bibnamefont{Menzel}}, \bibinfo {author}
  {\bibfnamefont{S.}~\bibnamefont{Saito}}, \bibinfo {author}
  {\bibfnamefont{K.}~\bibnamefont{Kakuyanagi}}, \bibinfo {author}
  {\bibfnamefont{H.}~\bibnamefont{Tanaka}}, \bibinfo {author}
  {\bibfnamefont{T.}~\bibnamefont{Meno}}, \bibinfo {author}
  {\bibfnamefont{K.}~\bibnamefont{Semba}}, \bibinfo {author}
  {\bibfnamefont{H.}~\bibnamefont{Takayanagi}},\ and\ \bibinfo {author}
  {\bibfnamefont{R.}~\bibnamefont{Gross}},\ }%
  \bibfield{journal}{%
  \bibinfo {journal} {Phys.\ Rev.\ B}\ }%
  \textbf{\bibinfo {volume} {76}},\ \bibinfo {pages} {214503} (\bibinfo {year}
  {2007})%
  \bibAnnoteFile{NoStop}{Deppe:2007a}%
\bibitem{Remark:1}%
  \BibitemOpen
  \bibinfo {note} {For our flux qubits we estimate a qubit-qubit coupling of
  $\SI{28}{\mega\hertz}$ due to a purely geometric coupling. In the region
  $\delta\Phi_{\rm x} \approx \pm 3~\text{m}\Phi_{\rm 0}$ where the one- and
  two-photon spectroscopy data of qubit 1 is recorded, the frequency separation
  of both qubits is more than \SI{2.5}{\giga\hertz}. Therefore, the presence of
  the second qubit can be safely disregarded.}%
  \bibAnnoteFile{Stop}{Remark:1}%
\bibitem{Schuster:2005a}%
  \BibitemOpen
  \bibfield{author}{%
  \bibinfo {author} {\bibfnamefont{D.~I.}\ \bibnamefont{Schuster}}, \bibinfo
  {author} {\bibfnamefont{A.}~\bibnamefont{Wallraff}}, \bibinfo {author}
  {\bibfnamefont{A.}~\bibnamefont{Blais}}, \bibinfo {author}
  {\bibfnamefont{L.}~\bibnamefont{Frunzio}}, \bibinfo {author}
  {\bibfnamefont{R.-S.}\ \bibnamefont{Huang}}, \bibinfo {author}
  {\bibfnamefont{J.}~\bibnamefont{Majer}}, \bibinfo {author}
  {\bibfnamefont{S.~M.}\ \bibnamefont{Girvin}},\ and\ \bibinfo {author}
  {\bibfnamefont{R.~J.}\ \bibnamefont{Schoelkopf}},\ }%
  \bibfield{journal}{%
  \bibinfo {journal} {Phys.\ Rev.\ Lett.}\ }%
  \textbf{\bibinfo {volume} {94}},\ \bibinfo {pages} {123602} (\bibinfo {year}
  {2005})%
  \bibAnnoteFile{NoStop}{Schuster:2005a}%
\bibitem{Simmonds:2004}%
  \BibitemOpen
  \bibfield{author}{%
  \bibinfo {author} {\bibfnamefont{R.~W.}\ \bibnamefont{Simmonds}}, \bibinfo
  {author} {\bibfnamefont{K.~M.}\ \bibnamefont{Lang}}, \bibinfo {author}
  {\bibfnamefont{D.~A.}\ \bibnamefont{Hite}}, \bibinfo {author}
  {\bibfnamefont{S.}~\bibnamefont{Nam}}, \bibinfo {author}
  {\bibfnamefont{D.~P.}\ \bibnamefont{Pappas}},\ and\ \bibinfo {author}
  {\bibfnamefont{J.~M.}\ \bibnamefont{Martinis}},\ }%
  \bibfield{journal}{%
  \bibinfo {journal} {Phys. Rev. Lett.}\ }%
  \textbf{\bibinfo {volume} {93}},\ \bibinfo {pages} {077003} (\bibinfo {year}
  {2004})%
  \bibAnnoteFile{NoStop}{Simmonds:2004}%
\bibitem{CohenTannoudji:2005}%
  \BibitemOpen
  \bibfield{author}{%
  \bibinfo {author} {\bibfnamefont{C.}~\bibnamefont{Cohen-Tannoudji}}, \bibinfo
  {author} {\bibfnamefont{B.}~\bibnamefont{Diu}},\ and\ \bibinfo {author}
  {\bibfnamefont{F.}~\bibnamefont{Lalo{\"e}}},\ }%
  \emph{\bibinfo {title} {Quantum Mechanics}}\ (\bibinfo {publisher}
  {Wiley-Interscience},\ \bibinfo {address} {New York},\ \bibinfo {year}
  {2005})%
  \bibAnnoteFile{NoStop}{CohenTannoudji:2005}%
\bibitem{Deppe:2009PhD}%
  \BibitemOpen
  \bibfield{author}{%
  \bibinfo {author} {\bibfnamefont{F.}~\bibnamefont{Deppe}},\ }%
  \emph{\bibinfo {title} {Superconducting Flux Quantum Circuits:
  Characterization, Quantum Coherence, and Controlled Symmetry Breaking}},\
  Ph.D. thesis,\ \bibinfo {school} {TU M{\"u}nchen} (\bibinfo {year} {2009})%
  \bibAnnoteFile{NoStop}{Deppe:2009PhD}%
\bibitem{Blais:2007}%
  \BibitemOpen
  \bibfield{author}{%
  \bibinfo {author} {\bibfnamefont{A.}~\bibnamefont{Blais}}, \bibinfo {author}
  {\bibfnamefont{J.}~\bibnamefont{Gambetta}}, \bibinfo {author}
  {\bibfnamefont{A.}~\bibnamefont{Wallraff}}, \bibinfo {author}
  {\bibfnamefont{D.~I.}\ \bibnamefont{Schuster}}, \bibinfo {author}
  {\bibfnamefont{S.~M.}\ \bibnamefont{Girvin}}, \bibinfo {author}
  {\bibfnamefont{M.~H.}\ \bibnamefont{Devoret}},\ and\ \bibinfo {author}
  {\bibfnamefont{R.~J.}\ \bibnamefont{Schoelkopf}},\ }%
  \bibfield{journal}{%
  \bibinfo {journal} {Phys. Rev. A}\ }%
  \textbf{\bibinfo {volume} {75}},\ \bibinfo {pages} {032329} (\bibinfo {year}
  {2007})%
  \bibAnnoteFile{NoStop}{Blais:2007}%
\bibitem{Sillanpaa:2007a}%
  \BibitemOpen
  \bibfield{author}{%
  \bibinfo {author} {\bibfnamefont{M.~A.}\ \bibnamefont{Sillanp{\"a}{\"a}}},
  \bibinfo {author} {\bibfnamefont{J.~I.}\ \bibnamefont{Park}},\ and\ \bibinfo
  {author} {\bibfnamefont{R.~W.}\ \bibnamefont{Simmonds}},\ }%
  \bibfield{journal}{%
  \bibinfo {journal} {Nature}\ }%
  \textbf{\bibinfo {volume} {449}},\ \bibinfo {pages} {438} (\bibinfo {year}
  {2007})%
  \bibAnnoteFile{NoStop}{Sillanpaa:2007a}%
\bibitem{Fedorov:2010}%
  \BibitemOpen
  \bibfield{author}{%
  \bibinfo {author} {\bibfnamefont{A.}~\bibnamefont{Fedorov}}, \bibinfo
  {author} {\bibfnamefont{A.~K.}\ \bibnamefont{Feofanov}}, \bibinfo {author}
  {\bibfnamefont{P.}~\bibnamefont{Macha}}, \bibinfo {author}
  {\bibfnamefont{P.}~\bibnamefont{Forn-D\'\i{}az}}, \bibinfo {author}
  {\bibfnamefont{C.~J. P.~M.}\ \bibnamefont{Harmans}},\ and\ \bibinfo {author}
  {\bibfnamefont{J.~E.}\ \bibnamefont{Mooij}},\ }%
  \bibfield{journal}{%
  \bibinfo {journal} {Phys. Rev. Lett.}\ }%
  \textbf{\bibinfo {volume} {105}},\ \bibinfo {pages} {060503} (\bibinfo {year}
  {2010})%
  \bibAnnoteFile{NoStop}{Fedorov:2010}%
\bibitem{Steffen:2010}%
  \BibitemOpen
  \bibfield{author}{%
  \bibinfo {author} {\bibfnamefont{M.}~\bibnamefont{Steffen}}, \bibinfo
  {author} {\bibfnamefont{S.}~\bibnamefont{Kumar}}, \bibinfo {author}
  {\bibfnamefont{D.~P.}\ \bibnamefont{DiVincenzo}}, \bibinfo {author}
  {\bibfnamefont{J.~R.}\ \bibnamefont{Rozen}}, \bibinfo {author}
  {\bibfnamefont{G.~A.}\ \bibnamefont{Keefe}}, \bibinfo {author}
  {\bibfnamefont{M.~B.}\ \bibnamefont{Rothwell}},\ and\ \bibinfo {author}
  {\bibfnamefont{M.~B.}\ \bibnamefont{Ketchen}},\ }%
  \bibfield{journal}{%
  \bibinfo {journal} {Phys. Rev. Lett.}\ }%
  \textbf{\bibinfo {volume} {105}},\ \bibinfo {pages} {100502} (\bibinfo {year}
  {2010})%
  \bibAnnoteFile{NoStop}{Steffen:2010}%
\end{thebibliography}

%

\end{document}